\pdfoutput=1
\documentclass[conference]{IEEEtran}

\usepackage{times}
\usepackage{epsfig}
\usepackage{subcaption}
\usepackage{amsmath}
\usepackage{amsfonts}
\usepackage{graphicx}
\usepackage{amssymb}
\usepackage{amstext}
\usepackage{latexsym}
\usepackage{color,soul}
\usepackage{ifthen}
\usepackage{enumerate}
\usepackage{graphicx}
\usepackage{multirow,soul}
\usepackage{verbatim}
\usepackage{array,tabularx}
\usepackage[nospace]{cite}
\usepackage[mathscr]{euscript}
\usepackage{tikz}
\usepackage{accents}
 \usepackage{cite}
 \usepackage{bm}
 \usepackage{arydshln}
\usepackage{hyperref}
\usepackage{amssymb,epsfig,graphics, graphicx, url,times,mathrsfs,algorithm,algorithmic,amsmath,array,latexsym,xspace,fancyhdr,wrapfig, xcolor,multirow,amsthm,cite}
\usepackage{caption}
\usepackage{amsmath}
\usepackage{soul}
\usepackage[inline]{enumitem}
\usepackage{arydshln}
\usepackage{enumitem}
\usepackage[makeroom]{cancel}
\usepackage{mathtools}
\usepackage{dblfloatfix}
\usepackage[utf8]{inputenc}

 \usepackage{tikz,pgfplots}
\usetikzlibrary{shapes.geometric}
\usetikzlibrary{shapes,snakes,patterns}
\usetikzlibrary{arrows,positioning,automata,calc}
\usetikzlibrary{plotmarks}
\tikzset{
    int/.style={
           rectangle,
           rounded corners,
           draw=black, thin, fill=black!20,
           minimum height=2em,
           inner sep=2pt,
           text centered,
           },
}
\definecolor{darkspringgreen}{rgb}{0.09, 0.45, 0.27}

\newcommand{\PreserveBackslash}[1]{\let\temp=\\#1\let\\=\temp}
\newcolumntype{C}[1]{>{\PreserveBackslash\centering}p{#1}}
\newcolumntype{R}[1]{>{\PreserveBackslash\raggedleft}p{#1}}
\newcolumntype{L}[1]{>{\PreserveBackslash\raggedright}p{#1}}

\newtheorem{theorem}{Theorem}

\newtheorem{claim}{Claim}
\newtheorem{example}{Example}
\newtheorem{definition}{Definition}
\newtheorem{corollary}{Corollary}
\newtheorem{remark}{Remark}

\IEEEoverridecommandlockouts

\pdfoutput=1

\begin{document}
\pdfoutput=1
\allowdisplaybreaks[1]
\newlength\figureheight
\newlength\figurewidth

\title{Codes for Correcting Localized Deletions}
\author{
\IEEEauthorblockN{Serge Kas Hanna\IEEEauthorrefmark{1},  Salim El Rouayheb\IEEEauthorrefmark{2} }

\IEEEauthorblockA{ 
\IEEEauthorrefmark{1} Institute for Communications Engineering, Technical University of Munich, Germany \\
\IEEEauthorrefmark{2} Department of Electrical and Computer Engineering, Rutgers University, USA  \\
Emails: serge.k.hanna@tum.de, salim.elrouayheb@rutgers.edu 
\thanks{This work was done while the first author was with the ECE Department, Rutgers University, USA.}
\thanks{This paper was presented in part at the 55th Annual Allerton Conference on Communication, Control, and Computing, 2017~\cite{Allerton}.}
\thanks{This work was supported in part by NSF Grant CCF 15-26875.}
} 
}
\maketitle

\begin{abstract} 
We consider the problem of constructing binary codes for correcting deletions that are localized within certain parts of the codeword that are unknown a priori. The model that we study is when $\delta \leq w$ deletions are localized in a window of size $w$ bits. These $\delta$ deletions do not necessarily occur in consecutive positions, but are restricted to the window of size $w$. The localized deletions model is a generalization of the bursty model, in which all the deleted bits are consecutive. In this paper, we construct new explicit codes for the localized model, based on the family of Guess \& Check codes which was previously introduced by the authors.  The codes that we construct can correct, with high probability, $\delta \leq w$ deletions that are localized in a single window of size~$w$, where $w$ grows with the block length. Moreover, these codes are systematic; have low redundancy; and have efficient deterministic encoding and decoding algorithms. We also generalize these codes to deletions that are localized within multiple windows in the codeword.
\end{abstract}
\section{Introduction}
Deletion and insertion errors are experienced in various communication and storage systems. These errors are often associated with various forms of loss of synchronization\mbox{\cite{M11,salimsync,Y14,V15}}. In many applications, the deletion and insertion errors tend to occur in bursts (consecutive errors), or are localized within certain parts of the information. This happens for example when synchronization is lost for a certain period of time, resulting in a series of errors within that time frame. Another example is in file synchronization, in applications such as  Dropbox, where large files are often edited by deleting and inserting characters in a relatively small part of the text (such as editing a paragraph), resulting in errors that are localized within that part of the file. Motivated by these applications, in this paper we focus on the problem of constructing codes that can correct localized deletions.

\subsection{Related work}
Deletion errors were first explored in the 1960s~\cite{VT65,L66,G61}. In 1966, Levenshtein~\cite{L66} showed that the codes constructed by Varshamov and Tenengolts (VT codes)~\cite{VT65} are capable of correcting a single deletion. Also in~\cite{L66}, Levenshtein derived non-constructive asymptotic bounds on the redundancy needed to correct deletions. The bounds showed that the redundancy needed to correct $\delta$ bit deletions in a codeword of length $n$ bits is asymptotically $c\delta \log n$, for some constant $c>0$. Moreover, for the case of a burst of exactly $\delta$ consecutive deletions, Levenshtein showed that at least $\log n +\delta-1$ redundant bits are required. Levenshtein's bounds were later improved by Cullina and Kiyavash~\cite{N14} for the general case of $\delta$ unrestricted deletions. Schoeny {\em et al.}~\cite{Sch17} applied similar techniques as in~\cite{N14} to derive a non-asymptotic lower bound for the case of a burst of $\delta$ deletions. The non-asymptotic bound derived in~\cite{Sch17} matched Levenshtein's asymptotic bound.

Several previous works studied the general problem of constructing binary codes that correct multiple unrestricted deletions~\mbox{($\delta>1$)}~\cite{H02,B16,G19,Vardy,GC,GCisit2017,Sima}. Intuitively, correcting a burst of two deletions is an easier problem compared to correcting two deletions that are not necessarily consecutive (unrestricted). This idea is also reflected through Levenshtein's bounds on the redundancy, which suggest that less redundancy would be required when the deletions occur in a burst. 

A separate line of work has focused on the problem of correcting deletions that occur in a {\em single} burst, i.e., consecutive deletions. Levenshtein~\cite{L67} constructed asymptotically optimal codes that can correct a burst of {\em at most} two deletions. \mbox{Cheng {\em et al.}~\cite{C14}} provided three constructions of codes that can correct a burst of {\em exactly} $\delta>2$ deletions. The lowest redundancy achieved by the codes in~\cite{C14} is $\delta\log(n/\delta+1)$. The fact that the number of deletions in the burst is exactly~$\delta$, as opposed to at most $\delta$, is a crucial factor in the code constructions in~\cite{C14}. 
Schoeny {\em et al.}~\cite{Sch17} proved the existence of codes that can correct a burst of exactly $\delta$ deletions and have at most \mbox{$\log n+(\delta-1)\log(\log n)+\delta-1$} redundancy, for sufficiently large $n$. The authors in~\cite{Sch17} also constructed codes that can correct a burst of at most $\delta$ deletions. Their results for the latter case improve on a previous result by Bours in~\cite{B94}. In the aforementioned results \cite{C14,Sch17,B94}, the size of the burst~$\delta$ is assumed to be a constant, i.e., $\delta$ does not grow with~$n$.
\begin{figure*}
\centering
\resizebox{1\textwidth}{!}{
\begin{tikzpicture}[node distance=2.5cm,auto,>=latex']
\draw (-0.6,0) -- (-0.45,0);
    \node [int] (c) [text width=0.8cm,align=center]{\scriptsize Binary to $q$-ary};
    \node (b) [left of=c,node distance=1.5cm, coordinate] {a};
    \node [int] (z) [right of=c, node distance=3.4cm,text width=2.75cm,align=center] {\scriptsize Systematic MDS $\left(k/\ell+c,k/\ell \right)$};
    \node [int] (y) [right of=z, node distance=3.8cm,text width=0.8cm,align=center] {\scriptsize $q$-ary to binary};
    \node [int] (y1) [right of=y, node distance=3cm, text width=2cm,align=center] {\scriptsize {Insert a buffer of $w+1$ bits between systematic and parity bits}};
    \node [coordinate] (end) [right of=c, node distance=2cm]{};
    \path[->] (b) edge node {\scriptsize $\mathbf{u}$} node[below] {\scriptsize $k$ bits} (-0.6,0);
    
    \draw[->] (c) edge node {\scriptsize $\mathbf{U}$} (z) ;
     \node(h1) [right of=c,node distance=1.2cm] [below] {\scriptsize $k/\ell$};
    \node(h2) [below of=h1,node distance=0.3cm] {\scriptsize symbols};
    \node(q1) [below of=c,node distance=0.9cm] {\scriptsize $q=2^{\ell}$};
    \node(q2) [below of=y,node distance=0.9cm] {\scriptsize $q=2^{\ell}$};
    \path[->] (z) edge node {\scriptsize $\mathbf{X}$} (y); 
    \node(i1) [right of=z,node distance=2.37cm] [below] {\scriptsize $k/\ell+c$};
   \node(name)[above of=i1,node distance=1.8cm] [above] {\scriptsize Guess \& Check (GC) codes};
    \node(i2) [below of=i1,node distance=0.3cm] {\scriptsize symbols};
    \node(e1) [right of=y1,node distance=3.3cm] {};
    \path[->] (y1) edge node {\scriptsize $\mathbf{x}$} (e1);
    \path[->] (y) edge node {} (y1);
    \node(t1) [right of=y,node distance=1.16cm] [below] {\scriptsize $k+c\ell$};
    \node(t2) [below of=t1,node distance=0.3cm] {\scriptsize bits};
    \node(tt1) [right of=y1,node distance=2.22cm] [below] {\scriptsize $k+c\ell+w+1$};
    \node(tt2) [below of=tt1,node distance=0.3cm] {\scriptsize bits}; 
    \node(b1) [above of=c,node distance=0.89cm] {\scriptsize Block I};
    \node(b2) [above of=z,node distance=0.665cm] {\scriptsize Block II};
    \node(b3) [above of=y,node distance=0.88cm] {\scriptsize Block III};
    \node(b4) [above of=y1,node distance=1.36cm] {\scriptsize Block IV};
    
    \draw[dashed] (-0.6,-1.3) rectangle (11.4,1.55);
\end{tikzpicture} }
\captionsetup{font=footnotesize}
\caption{Encoding block diagram of GC codes for correcting $\delta\leq w$ deletions that are localized within a single window of size at most $w$ bits.  Block~I:~The binary message of length $k$ bits is chunked into adjacent blocks of length $\ell$ bits each, and each block is mapped to its corresponding symbol in $GF(q)$ where $q=2^{\ell}$. If the last block after chunking contains fewer than $\ell$ bits, it is assumed to be padded with zeros. Block~II:~The resulting string is encoded using a systematic $\left(k/\ell+c,k/\ell \right)$ $q$-ary MDS code where $c$ is the number of parity symbols and $q=k>k/\ell+c$. Block~III:~The symbols in $GF(q)$ are mapped to their binary representations. Block~IV:~A buffer of $w$ zeros followed by a single one is inserted between the systematic and the parity bits. \vspace{-0.4cm}}
\label{fig:1}
\end{figure*}

\subsection{Contributions}
In this paper, we design efficient codes for a more general model, where the $\delta$ deletions are not necessarily consecutive but are localized in a window of size at most $w$ bits. We also generalize these codes to the case of deletions that are localized within {\em multiple} windows. To the best of our knowledge, the only previous work on codes for localized deletions is the one by Schoeny {\em et al.} in~\cite{Sch17} where the authors constructed codes for the particular case of a single window with $w=3$ or $w=4$. Moreover, the case of multiple windows has not been studied in the literature even for the simpler model of bursty deletions. 

The codes that we present are based on our previous work in~\cite{GC}, where we proposed the Guess \& Check (GC) codes\footnote{An implementation of GC codes can be found and tested online on the link in~\cite{Software}.} for the general problem of correcting \mbox{$\delta> 1$} unrestricted deletions. In this work, we exploit the localized nature of the deletions to modify the schemes in~\cite{GC} and obtain codes that have an asymptotically optimal rate, and efficient encoding and decoding algorithms, for a number of localized deletions that {\em grows} with the block length. Furthermore, for a uniform iid message, and for arbitrary deletion positions that are chosen independently of the codeword, these codes have an asymptotically vanishing probability of decoding failure\footnote{The term {\em decoding failure} means that the decoder cannot make a correct decision and outputs a ``failure to decode" error message.}. Note that previous works on bursty deletions (e.g., \cite{Sch17,C14,B94}), have focused on constructing {\em zero-error} codes that can correct a {\em constant} number of consecutive deletions ($\delta$ does not grow with $n$), which corresponds to correcting worst-case bursts of constant size. However, it remains open to show whether the zero-error based constructions in \cite{Sch17,C14,B94} can result in practical polynomial time encoding and decoding schemes for the problem.  

Namely, our contributions are the following: \begin{enumerate*}[label=(\roman*)] \item We construct new explicit codes, that can correct, with high probability, $\delta\leq w$ deletions that are localized within a single window of size \mbox{$w$} bits. In our construction, we allow $\delta$ and $w$ to grow with the blocklength. \item These codes have a redundancy that is lower than the codes existing in the literature on bursts of deletions.  Also, the code rate of these codes is asymptotically optimal for a sublinear number of localized deletions. \item The complexity of the encoding algorithm is near-linear, and the decoding complexity is quadratic.  \item We provide numerical simulations on the performance of these codes, and compare the simulation results to our theoretical bound on the probability of decoding failure. \item We generalize our approach to obtain codes that can correct deletions that are localized within multiple windows in the codeword. \end{enumerate*}

\subsection{Organization} 
The paper is organized as follows. In Section~\ref{sec:2}, we formally present the model for localized deletions and introduce the basic notation and terminology used throughout the paper.  We state and discuss our main results in Section~\ref{sec:3}, and compare them to the most recent works on unrestricted and bursty deletions. In Section~\ref{sec:4}, we provide an encoding and decoding example of GC codes for correcting deletions thats are localized within a single window. In Section~\ref{sec:5}, we describe in detail our encoding and decoding schemes and discuss the choices of the code parameters. The results of the numerical simulations are presented in Section~\ref{sec:6}. In Section~\ref{sec:7}, we show the generalization of our codes to the case of deletions that are localized within multiple windows. The proof of the main result of this paper is given in Section~\ref{sec:8}.

\section{Preliminaries}
\label{sec:2}
\noindent In this paper, we consider the following models for deletions:
\begin{enumerate}
\item Bursts of fixed lengths: the deletions occur in $z\geq 1$ bursts, where each burst corresponds to exactly $\delta\geq 1$ consecutive deletions. 
\item Bursts of variable lengths: the deletions occur in \mbox{$z\geq 1$} bursts, each of length at most $\delta\geq 1$ bits.
\item Localized deletions: the deletions are restricted to $z\geq 1$ windows. The size of each window is at most $w$ bits, and the number of deletions within each window is at most $\delta \leq w$. The bits that are deleted in a certain window are not necessarily consecutive.
\begin{example}[Localized Deletions]
Let $z=2,w=4,$ and $\delta=3$. Let $\mathbf{x}$ be the transmitted string and $\mathbf{y}$ be the received string. The deletions (underlined) are localized in $z=2$ windows.
\begin{align*}
\mathbf{x}&=1~0~\overbracket{\underline{0}~\underline{1}~\color{black}0~\underline{1}}^{\text{window~1}}~\color{black}0~0~1~0~1~0~0~1~\overbracket{\underline{0}~\color{black}0~\underline{1}~\color{black}0}^{\text{window~2}}~1~1~0, \\
\mathbf{y}&=1~0~0~0~0~1~0~1~0~0~1~0~0~1~1~0.
\end{align*}
\end{example}
Note that the two previously mentioned bursty models correspond to special cases of the localized model. Hence, a code that can correct localized deletions can also correct bursts of fixed or variable lengths. 
\item Unrestricted deletions: at most $\delta\geq 1$ deletions can occur anywhere in the transmitted string.
\end{enumerate}
We denote by $k$ and $n$ the lengths in bits of the message and codeword, respectively. The regime we study in this paper is when $w=o(k)$. The encoding block diagram of GC codes for correcting deletions that are localized within a single window $(z=1)$ is shown in Fig.~\ref{fig:1}. We denote binary and $q$-ary vectors by lower and upper case bold letters respectively, and random variables by calligraphic letters.  All logarithms in this paper are of base~$2$. 

For the aforementioned deletion models, we study the setting where the positions of the deletions are arbitrary but are chosen independently of the codeword. We also assume that the information message is uniform iid. We denote by $F$ the event that corresponds to a decoding failure, i.e., when the decoder cannot make a correct decision and outputs a ``failure to decode" error message. $Pr(F)$ is the probability of decoding failure where this probability is for a uniform iid message and for arbitrary deletion positions that are chosen independently of the codeword.

The notations used in this paper are summarized in Table~I.
\begin{table}[h]
\centering
\setlength\extrarowheight{1.6pt}
 \begin{tabular}{|c|l|c|l|}
\hline
Variable & Description \\ \hline
$\mathbf{u}$ & message  \\ \hline
$k$ & length of the message in bits  \\ \hline
$\mathbf{x}$ & codeword  \\ \hline
$n$ & length of codeword in bits   \\ \hline
$w$ & size of the deletion window \\ \hline
$\delta$ & number of deletions  \\ \hline
 $c$ & number of MDS parity symbols \\ \hline
$\ell$ & chunking length used for encoding (Fig.~\ref{fig:1}) \\ \hline
$q$ & field size given by $2^{\ell}$ \\ \hline
\end{tabular}
\captionsetup{font=footnotesize}
\caption{\footnotesize{Summary of the notations used in the paper.}}
\vspace{-0.3cm}
\label{t}
\end{table}

\section{Main Results}
In this section, we state our two main results in this paper and compare them to the most recent works on other deletion models. Theorem~\ref{thm:1} and Theorem~\ref{thm:2} cover the case of deletions that are localized in a single window ($z=1$), with $w=o(\log k)$ and $w=\Omega(\log k)$, respectively. We discuss the generalization of these theorems to $z>1$ windows in Section~\ref{sec:7}.
\label{sec:3}

\begin{theorem}[Single window with $w=o(\log k)$]
\label{thm:1} 
Guess \&~Check (GC) codes can correct in polynomial time $\delta\leq w$ deletions that are localized within a single window of size at most $w$ bits, where \mbox{$w=o(\log k)$}. The code has the following properties:
\begin{enumerate}[leftmargin=*]
\item Redundancy: $n-k = (c+1)\log k+1$ bits,
\item Encoding complexity is $\mathcal{O}(k \log k)$, and  decoding complexity is $\mathcal{O}\left(k^2 \right)$,
\item Probability of decoding failure: 
$Pr(F)\leq k^{-(c-4)}/\log k, $
\end{enumerate}
where $c$ is a code parameter that represents the number of parity symbols.
\end{theorem}

\begin{theorem} [Single window with $w=\Omega(\log k)$]
\label{thm:2}
Guess \&~Check (GC) codes can correct in polynomial time $\delta\leq w$ deletions that are localized within a single window of size at most $w$ bits, where \mbox{$w=\Omega(\log k)$} and $w= o(k)$. The code has the following properties:
\begin{enumerate}[leftmargin=*]
\item Redundancy: $n-k = (c+1)w+1$ bits,
\item Encoding complexity is $\mathcal{O}(kw)$, and  decoding complexity is $\mathcal{O}\left(k^2\right)$,
\item Probability of decoding failure: 
$Pr(F)\leq (k/w) 2^{-w(c-3)}$,
\end{enumerate}
where $c$ is a code parameter that represents the number of parity symbols.
\end{theorem}

\begin{corollary}
For both regimes $w=o(\log k)$ and $w=\Omega(\log k)$ described in Theorems~\ref{thm:1} and \ref{thm:2}, the probability of decoding failure of GC codes for a single window of localized deletions vanishes asymptotically in $k$ for $c\geq 4$.
\end{corollary}

The proofs of Theorems~\ref{thm:1} and~\ref{thm:2} are provided in Section~\ref{sec:8}. The proof of Corollary~1 follows directly from replacing the value of $c$ is the expression of the probability of decoding failure in Theorems~\ref{thm:1} and~\ref{thm:2} and is therefore omitted. The code properties in Theorems~\ref{thm:1} and~\ref{thm:2} show that: 

\begin{enumerate}[leftmargin=*] 
\item The redundancy is logarithmic in $k$ for $w=o(\log k)$; and linear in $w$ for $w=\Omega(\log k)$. One can easily verify that this corresponds to a  code rate $R=k/n$ that is asymptotically optimal for both regimes, i.e., $R\to 1$ as $k\to +\infty$.
 \item The decoding complexity is quadratic; and the encoding complexity is near-linear for small $w$, and subquadratic for large $w$. 
 \item The probability of decoding failure vanishes asymptotically in $k$ for $c\geq 4$. Moreover, this probability decreases exponentially in $c$ for a fixed $k$. 
 \end{enumerate}

In terms of the code construction, the only difference between the two regimes: $w=o(\log k)$ and $w=\Omega(\log k)$, is the choice of the value of the chunking length $\ell$ (Fig.~\ref{fig:1}). For $w=\Omega(\log k)$ we set $\ell=w$ (Theorem~\ref{thm:2}), whereas for $w=o(\log k)$ we set $\ell=\log k$ (Theorem~\ref{thm:1}). The reason we differentiate between these two regimes is because our analysis in Section~\ref{sec:8} shows that $\ell=\Omega(\log k)$ is always required to guarantee that GC codes have an asymptotically vanishing probability of decoding failure. So although the choice of $\ell=w$ gives a good trade-off between redundancy and decoding complexity as we explain in Section~\ref{sec:5}, it does not guarantee a low probability of error when $w=o(\log k)$.

\begin{remark}
The bounds on the probability of decoding failure in Theorems~\ref{thm:1} and~\ref{thm:2} hold for any window position, and any $\delta\leq w$ deletion positions within this window, that are chosen independently of the codeword. Hence, the same result can be also obtained for any random distribution on the positions of the window and the deletions (like the uniform distribution for example), by applying the law of total probability. 
\end{remark}

{\em Comparison to recent work:}
As previously mentioned, there has been many works in the literature on other deletion models such as the unrestricted and the bursty models. Next, we compare our results on localized deletions (Theorems~\ref{thm:1} and \ref{thm:2}) to the most recent works on these models to provide helpful context to the reader. 

Recently, Sima and Bruck \cite{Sima} introduced optimal codes that can correct $\delta$ unrestricted deletions with zero-error, where $\delta$ is a constant that is fixed with respect to the blocklength. The redundancy of these codes is $\mathcal{O}(\delta \log n)$, which is optimal when compared to Levenshtein's bound. The encoding/decoding complexity is $\mathcal{O}(n^{2\delta+1})$, which is polynomial in $n$ and exponential in $\delta$. 

The work that is closest to our work in spirit is the one by Schoeny {\em et al.}~\cite{Sch17} where the authors introduce codes that can correct a burst of $\delta$ deletions with zero-error, where $\delta$ is a constant that is fixed with respect to the blocklength. For the case of bursts of fixed lengths (exactly $\delta$ deletions), the authors prove the existence of codes that have at most \mbox{$\log n+(\delta-1)\log(\log n)+\delta-1$} redundancy, for sufficiently large $n$. As for the case of bursts of variable lengths (at most $\delta$ deletions), the redundancy is at most \mbox{$(\delta-1)\log n+\big(\binom{\delta}{2}-1\big)\log (\log n)+\binom{\delta}{2}+\log(\log \delta)$}. The authors in~\cite{Sch17} do not provide an explicit polynomial time encoding algorithm for these codes.

Comparing our work on localized deletions (Theorems~\ref{thm:1} and \ref{thm:2}) to the works mentioned above, we can observe the following. 
\begin{enumerate}
\item The number of deletions in both~\cite{Sima} and \cite{Sch17} is constant, whereas in our work the number of localized deletions $w=o(k)$ grows with the message length.

\item In the regime where the number of the deletions is constant, the redundancy in our construction is lower than the redundancy in~\cite{Sima}, which is intuitive since the deletions in the localized model are restricted. Furthermore, in the same regime, the redundancy of our codes is lower than the redundancy in~\cite{Sch17} for variable bursts, and can also be lower than the redundancy for fixed bursts for certain values of $\delta$ and $n$. 

\item Our codes are explicit, where the encoding complexity is near-linear and the decoding complexity is quadratic. These complexities are much lower than that of the codes in~\cite{Sima}. Furthermore, as previously mentioned, the codes in~\cite{Sch17} do not have an explicit encoding algorithm.

\item The gains provided by our codes come at the expense of a small probability of decoding failure that vanishes asymptotically. The codes in~\cite{Sima} and~\cite{Sch17} have zero-error.

\end{enumerate}

\section{Examples} 
\label{sec:4}
In this section, we provide encoding and decoding examples of GC codes for correcting $\delta\leq w$ deletions that are localized within a single window of size $w=\log k$ bits. The chunking length (Fig.~\ref{fig:1}) is set to $\ell=w=\log k$.  
\begin{example}[Encoding]
\label{ex:1} Consider a message $\mathbf{u}$ of length $k=16$ given by $\mathbf{u}=1100101001111000$.
$\mathbf{u}$ is encoded by following the different encoding blocks illustrated in Fig.~\ref{fig:1}. \\
\noindent $1)$ {\em Binary to $q$-ary (Block I, Fig.~\ref{fig:1})}. The message $\mathbf{u}$ is chunked into adjacent blocks of length $\ell=\log k=4$ bits each, 
\begin{equation*}
\mathbf{u}=\underbrace{\overbracket{1~1~0~0}^{\text{block 1}}}_{\alpha^{6}}~\underbrace{\overbracket{1~0~1~0}^{\text{block 2}}}_{\alpha^{9}}~\underbrace{\overbracket{0~1~1~1}^{\text{block 3}}}_{\alpha^{10}}~\underbrace{\overbracket{1~0~0~0}^{\text{block 4}}}_{\alpha^{3}}\color{black}.
\end{equation*}
Each block is then mapped to its corresponding symbol in $GF(q)$, $q=2^{\ell}=2^4=16$. This results in a string $\mathbf{U}$ which consists of $k/\log k=4$ symbols in $GF(16)$. The extension field used here has a primitive element $\alpha$, with $\alpha^4=\alpha+1$.  Hence, 
we obtain $\mathbf{U}=(\alpha^{6},\alpha^{9},\alpha^{10},\alpha^3\color{black})\in GF(16)^4$. \\
$2)$ {\em Systematic MDS code (Block II, Fig.~\ref{fig:1})}. $\mathbf{U}$ is then encoded using a systematic $(k/\log k +c,k/\log k)=(7,4)$ MDS code over $GF(16)$, with $c=3$. The encoded string is denoted by $\mathbf{X}\in GF(16)^3$ and is given by multiplying $\mathbf{U}$ by the following code generator matrix\footnote{The MDS generator matrix used in this example is based on a Vandermonde matrix. For the general case we construct the generator matrix by concatenating an identity matrix with a Cauchy matrix.}
\begin{align*}
\mathbf{X} &= \left(\alpha^{6},\alpha^9,\alpha^{10},\alpha^3\color{black}\right)
\left(\
\begin{matrix}  
 1 &0 &0 &  0 &  1 &  1 &  1\\
 0 &  1 &  0 &  0 &  1 &  \alpha &  \alpha^2\\
 0 &  0 &  1 &  0 &  1 &  \alpha^2 &  \alpha^4\\
 0 &  0 &  0 &  1 &  1 &  \alpha^3 &  \alpha^6 \\
\end{matrix} \right) \\
&= (  \alpha^{6},\alpha^9,\alpha^{10},\alpha^3,  \alpha^{14},\alpha^{3},\alpha \color{black}).
\end{align*}
$3)$ {\em $q$-ary to binary (Block III, Fig.~\ref{fig:1})}. The binary representation of $\mathbf{X}$, of length $n=k+3\log k=28$ bits, is $ 1100~1010~0111~1000~ 1001~ 1000~ 0001\color{black}$. \\
$4)$ {\em Adding a buffer of $w+1$ bits (Block IV, Fig.~\ref{fig:1})}. A buffer of $w=\log k=4$ zeros followed by single one is inserted between the systematic and parity bits. 
The binary codeword to be transmitted is of length $33$ bits and is given by
\begin{equation*}
\mathbf{x}= 1100~1010~0111~1000~\color{black}\overbracket{\mathbf{00001}}^{\text{buffer}}~ 1001~ 1000~ 0001\color{black}.
\end{equation*}
\end{example}
\vspace{-0.5cm}
Now we explain the high-level idea of decoding. Since \mbox{$\delta\leq w$} and \mbox{$w=\ell=\log k$}, then the $\delta$ deletions can affect at most two adjacent blocks in $\mathbf{x}$. The goal of the decoder is to recover the systematic part of $\mathbf{x}$, so it considers following guesses:  1)~blocks~$1$~and~$2$ are affected by the deletions; 2)~blocks~$2$~and~$3$ are affected; 3)~blocks~$3$~and~$4$ are affected. For each of these guesses the decoder: \begin{enumerate*}[label=(\roman*)] \item Chunks the received sequence based on its guess on the locations of the two affected blocks. \item Considers the two affected blocks erased and decodes them by using the first two MDS parity symbols. \item Checks whether the decoded string is consistent with the third MDS parity and with the received sequence. \end{enumerate*}

\begin{example}[Decoding] \label{ex:1a}
Suppose that the $7^{th}$, $9^{th}$ and $10^{th}$ bit of $\mathbf{x}$ are deleted. Hence, the decoder receives the following $30$ bit string $\mathbf{y}$,
\begin{equation*}
\mathbf{y}= 1100100111000\color{black}\mathbf{0}\mathbf{0}\mathbf{0}\mathbf{0}\mathbf{1} 100110000001\color{black}.
\end{equation*}
Note that the window size $w$ is known at the decoder, and the number of deletions $\delta$ can be determined by the difference between $n$ (code parameter) and the length of the received string $\mathbf{y}$. In this example, we have $w=\log k=4$ and $\delta=3$. Moreover, the localized $\delta\leq w$ deletions cannot affect both the systematic and the parity bits simultaneously, since these two are separated by a buffer of size $w+1$ bits.  
Therefore, we consider the following two scenarios:  \begin{enumerate*}[label=(\roman*)] \item If the deletions affected the parity bits, then the decoder simply outputs the systematic bits\footnote{The information is in the systematic bits, hence the decoder does need to recover parity bits.}. \item If the deletions affected the systematic bits, then the decoder goes over the guesses as explained previously. \end{enumerate*}

The buffer is what allows the decoder to determine which of the two previous scenarios to consider. The decoder observes the \mbox{$(k+w-\delta+1)^{th}=18^{th}$} bit in $\mathbf{y}$ (underlined), 
\begin{equation*}
 1100100111000\color{black}\mathbf{0}\mathbf{0}\mathbf{0}\mathbf{0}\textbf{\underline{$\mathbf{1}$}} 100110000001\color{black}.
\end{equation*}
Based on the value of the observed bit, the decoder can determine whether the deletions affected the systematic bits or not. The previous operation can be done with zero-error, we explain it in more detail in Section~\ref{sec:5}.
In this example, the fact that the observed bit is a one indicates that the one in the buffer has shifted $\delta$ positions to the left. Hence, the decoder considers that the deletions have affected the systematic bits, and thus proceeds with making its guesses as we explain next. Henceforth, the buffer is removed from the string.

The decoder goes through all the possible \mbox{$k/\log k-1=3$} cases (guesses), where in each case $i$, $i=1,\ldots,3$, the deletions are assumed to have affected blocks $i$ and $i+1$, and $\mathbf{y}$ is chunked accordingly. Given this assumption, symbols $i$ and $i+1$ are considered erased and erasure decoding is applied over $GF(16)$ to recover these two symbols. Without loss of generality, we assume that the first two parities $p_1=\alpha^{14} $ and $p_2=\alpha^3$ are used for decoding the two erasures. The decoded $q$-ary string in case $i$ is denoted by $\mathbf{Y_i}\in GF(16)^4$, and its binary representation is denoted by $\mathbf{y_i}\in GF(2)^{16}$. The three cases are shown below: \\
{\em \underline{Case 1}:} The deletions are assumed to have affected blocks~1~and~2. Hence, $\mathbf{y}$ is chunked as follows \begin{equation*}
  \underbrace{1~1~0~0~1}_{\mathcal{E}}~\underbrace{0~0~1~1}_{\alpha^4}~\underbrace{1~0~0~0}_{\alpha^{3}}~ \underbrace{1~0~0~1}_{\alpha^{14}}~\underbrace{1~0~0~0}_{\alpha^3}~\underbrace{0~0~0~1}_{1}\color{black},
\end{equation*}
where $\mathcal{E}$ denotes the bits corresponding to symbols 1 and 2 that are considered to be erased. Applying erasure decoding over $GF(16)$, the recovered values of symbols 1 and 2 are $\alpha^{2}$ and $\alpha^{5}$, respectively. Hence, the decoded $q$-ary string \mbox{$\mathbf{Y_1} \in GF(16)^4$} is
\begin{equation*}
\mathbf{Y_1}=( \alpha^{2},\alpha^{5},\alpha^{4},\alpha^{3}\color{black}) .
\end{equation*}
Its equivalent in binary $\mathbf{y_1}\in GF(2)^{16}$ is
\begin{equation*}
\mathbf{y_1} =  \underbrace{0~1~0~0}_{\alpha^{2}}~\underbrace{0~1~1~0}_{\alpha^{5}}~\underbrace{0~0~1~1}_{\alpha^{4}}~\underbrace{1~0~0~0}_{\alpha^{3}}\color{black}.
\end{equation*}
Notice that the concatenated binary representation of the two decoded erasures $(01000110)$, is not a supersequence of the sub-block $(11001)$, which was denoted by $\mathcal{E}$. Hence, the decoder can immediately point out that the assumption in this case is wrong, i.e., the deletions did not affect blocks~1~and~2. Throughout the paper we refer to such cases as {\em impossible} cases. Another way for the decoder to check whether this case is possible is to test if $\mathbf{Y_1}$ is consistent with the third parity $p_3=1$. However, the computed parity is 
\begin{equation*}
\left( \alpha^{2},\alpha^{5},\alpha^{4},\alpha^{3}\color{black}\right)\left(1,\alpha^2,\alpha^{4},\alpha^{6}\right)^{T}=0 \neq   1 \color{black}.
\end{equation*}
Therefore, this is an additional reason which shows that case~1 is {\em impossible}. \\
{\em \underline{Case 2}:}  The deletions are assumed to have affected blocks 2 and 3, so the sequence is chunked as follows
\begin{equation*}
  \underbrace{1~1~0~0}_{\alpha^{6}}~\underbrace{1~0~0~1~1}_{\mathcal{E}}~\underbrace{1~0~0~0}_{\alpha^{3}}~ \underbrace{1~0~0~1}_{\alpha^{14}}~\underbrace{1~0~0~0}_{\alpha^{3}}~\underbrace{0~0~0~1}_{1}\color{black}.
\end{equation*}
Applying erasure decoding, the recovered values of symbols~2~and~3 are $\alpha^9$ and $\alpha^{10}$, respectively. The decoded binary string is
\begin{equation*}
\mathbf{y_2} =  \underbrace{1~1~0~0}_{\alpha^{6}}~\underbrace{1~0~1~0}_{\alpha^9}~\underbrace{0~1~1~1}_{\alpha^{10}}~\underbrace{1~0~0~0}_{\alpha^{3}}~\color{black}.
\end{equation*}
In this case, the concatenated binary representation of the two decoded erasures $(10100111)$ is a supersequence of the sub-block $(10011)$. Moreover, it is easy to verify that the decoded string is consistent with the third parity $p_3=1$. Therefore, we say that case~2 is {\em possible}. \\
\noindent {\em \underline{Case 3}:}  The deletions are assumed to have affected blocks 3 and 4, so the sequence is chunked as follows
\begin{equation*}
  \underbrace{1~1~0~0}_{\alpha^{6}}~\underbrace{1~0~0~1}_{\alpha^{14}}~\underbrace{1~1~0~0~0}_{\mathcal{E}}~ \underbrace{1~0~0~1}_{\alpha^{14}}~\underbrace{1~0~0~0}_{\alpha^{3}}~\underbrace{0~0~0~1}_{1}\color{black}.
\end{equation*}
The decoded binary string is
\begin{equation*}
\mathbf{y_3} =  \underbrace{1~1~0~0}_{\alpha^{6}}~\underbrace{1~0~0~1}_{\alpha^{14}}~\underbrace{1~1~0~0}_{\alpha^{6}}~\underbrace{0~0~0~0}_{0}~\color{black}.
\end{equation*}
In this case, the concatenated binary representation of the two decoded erasures $(11000000)$ is a supersequence of the sub-block $(11000)$. However, it is easy to verify that the decoded string is not consistent with $p_3=1$. Therefore, case~3 is {\em impossible}.

After going through all the cases, case~2 stands alone as the only {\em possible} case. So the decoder declares successful decoding and outputs $\mathbf{y_2}$ ($\mathbf{y_2}=\mathbf{u}$). 
\end{example}
\begin{remark}
Sometimes the decoder may find more than one {\em possible} case resulting in different decoded strings. In that situation, the decoder cannot know which of the cases is the correct one, so it declares a decoding failure. Although a decoding failure may occur, Theorems~\ref{thm:1} and~\ref{thm:2} indicate that its probability vanishes as length of the message $k$ goes to infinity.
\end{remark}

\section{Encoding and Decoding for Localized Deletions}
\label{sec:5}
As previously mentioned, our schemes for correcting localized deletions extend from the encoding and decoding schemes of Guess \& Check (GC) codes in~\cite{GC}, which are designed for correcting $\delta$ deletions that are not necessarily localized (i.e., unrestricted). In this section, we give an overview of the encoding and decoding schemes in~\cite{GC}, and explain how we exploit the localized nature of the deletions to modify these schemes and obtain codes having the properties shown in Section~\ref{sec:3}. We first start by restating the main result in~\cite{GC}. 
\begin{theorem} 
\label{thm:5}(\hspace{-0.03cm}\cite{GC})
Guess \& Check (GC) codes can correct in polynomial time up to a constant number of $\delta$ deletions. Let $c>\delta$ be a constant integer. The code has the following properties:
\begin{enumerate}[leftmargin=*]
\item Redundancy: $n-k = c(\delta+1) \ell$ bits. 
\item Encoding complexity is $\mathcal{O}(k \ell)$, and  decoding complexity is $\mathcal{O}\left(k^{\delta+1}/\ell^{\delta-1}\right)$.
\item Probability of decoding failure: $$Pr(F) = \mathcal{O} \big( (k/\ell)^{\delta}2^{-\ell(c-\delta)}\big).$$
\end{enumerate}
\end{theorem}
Next, we explain the encoding and decoding using GC codes for the following two models: (1)~Unrestricted deletions; and (2)~Localized deletions in a single window ($z=1$).
\subsection{Encoding using GC codes} 
The first three encoding blocks are the same as the ones shown in Fig.~\ref{fig:1} for both models mentioned above. For unrestricted deletions, the choice of the chunking length $\ell$ presents a trade-off between redundancy, complexity, and probability of decoding failure. For localized deletions, we specify the value of $\ell$ based on the size of the window $w$. Namely,
\begin{equation}
\label{eqell}
  \ell = \left\{\def\arraystretch{1.2}%
  \begin{array}{@{}c@{\quad}l@{}}
    \log k & \text{if $w=o(\log k)$,}\\
    w & \text{if $w=\Omega(\log k)$ and $w=o(k)$}.\\
  \end{array}\right.
\end{equation}
\noindent As for the last encoding block (Block~IV, Fig.~\ref{fig:1}):
\noindent \subsubsection{Unrestricted deletions} The parity bits are encoded using a $(\delta+1)$ repetition code, where each parity bit is repeated $(\delta+1)$ times. The repetition code protects the parities against any deletions, and allows them to be recovered at the decoder.
\noindent \subsubsection{Localized deletions in a single window ($z=1$)} The systematic and parity bits are separated by a buffer of size $w+1$ bits, which consists of $w$ zeros followed by a single one. In the upcoming decoding section, we explain how the decoder uses this buffer to detect whether the deletions affected the systematic bits or not.

\subsection{Decoding using GC codes} 
\noindent \subsubsection{Unrestricted deletions~\cite{GC}}
The approach presented in~\cite{GC} for decoding up to $\delta$ unrestricted deletions is the following: \\
\noindent $(a)$ Decoding the parity bits: the decoder recovers the parity bits which are protected by a $(\delta+1)$ repetition code.\\
\noindent $(b)$ The guessing part: the number of possible ways to distribute the $\delta$ deletions among the $k/\ell$ blocks is $t~=~\binom{k/\ell+\delta-1}{\delta}.$ These possibilities are indexed by \mbox{$i, i=1,\ldots,t,$} and each possibility is referred to by case~$i$.\\
\noindent $(c)$ The checking part: for each case $i$, $i=1,\ldots,t$, the decoder:  \begin{enumerate*}[label=(\roman*)] \item Chunks the sequence based on the corresponding assumption on the locations of the $\delta$ deletions. \item Considers the affected blocks erased and maps the remaining blocks to their corresponding symbols in $GF(q)$. \item Decodes the erasures using the first $\delta$ parity symbols. \item Checks whether the case is {\em possible} or not by testing if the decoded string is consistent with the received string and with the last $c-\delta$ parity symbols. \end{enumerate*} The criteria used to check if a case is possible or not are given in Definition~\ref{def:1}.
\begin{definition} \label{def:1} For $\delta$ deletions, a case $i$, $i=1,\ldots,t$, is said to be {\em possible} if it satisfies the following two criteria simultaneously. Criterion 1: the decoded $q$-ary string in case~$i$ is consistent with the last $c-\delta$ parities simultaneously. Criterion~2: the binary representations of all the decoded erasures are supersequences of their corresponding sub-blocks.
\end{definition}

\subsubsection{Localized deletions in a single window ($z=1$)} Consider the case of decoding \mbox{$\delta\leq w$} deletions that are localized within a single window of size at most $w$ bits. Since $w\leq \ell$ from~\eqref{eqell}, then the deletions can affect at most two adjacent blocks in the codeword. Therefore, in terms of the GC code construction, correcting the $\delta\leq w$ deletions corresponds to decoding at most two block erasures.  Hence, the localized nature of the deletions enables the following simplifications to the scheme:  \begin{enumerate*}[label=(\roman*)] \item The total number of cases to be checked by the decoder is reduced to $k/\ell-1$ since at most two adjacent blocks (out of $k/\ell$ blocks) can be affected by the deletions. \item Instead of protecting the parity bits by a $(\delta+1)$ repetition code, it is sufficient to separate them from the systematic bits by inserting a buffer of size $w+1$ bits, composed of $w$ zeros followed by a single one. \end{enumerate*}

Now we explain the decoding steps. Note that the~$\delta\leq w$ localized deletions cannot affect the systematic and the parity bits simultaneously since these two are separated by a buffer of size $w+1$ bits. The decoder uses this buffer to detect whether the deletions have affected the systematic bits or not. The buffer is composed of $w$ zeros followed by a single one, and its position ranges from the \mbox{$(k+1)^{th}$}~bit to the \mbox{$(k+w+1)^{th}$}~bit of the transmitted string. Let $\mathbf{y}_{\lambda}$ be the bit in position $\lambda$ in the received string, where \mbox{$\lambda\triangleq k+w-\delta+1$}. 
The decoder observes $\mathbf{y}_{\lambda}$ \footnote{The decoder knows the values of $k,n,$ and $w$ (code parameters), and can determine the value of $\delta$ by calculating the difference between $n$ and the length of the received string.}: (1)~If \mbox{$\mathbf{y}_{\lambda}=1$}, then this means that the one in the buffer has shifted $\delta$ positions to the left because of the deletions, i.e., all the deletions occurred to the left of the one in the buffer. In this case, the decoder considers that the deletions affected the systematic bits, and therefore proceeds to the guessing and checking part. It applies the same steps as in the case of $\delta$ deletions, while considering a total of \mbox{$k/\ell-1$} cases, each corresponding to two adjacent block erasures. In each case, the last $c\ell$ bits of the received string (parities) are used to decode the first $k-\delta$ bits. (2)~If $\mathbf{y}_{\lambda}=0$, then this indicates that the $\delta$ deletions occurred to the right of the first zero in the buffer, i.e., the systematic bits were unaffected. In this case, the decoder simply outputs the first $k$ bits of the received string. 
\subsection{Discussion}

\subsubsection{Choice of the chunking length $\ell$ for localized deletions} Recall that the choice of the value of $\ell$ affects the redundancy of the code which is given by $c\ell + w + 1$ (Fig.\ref{fig:1}). For this reason, we would like keep $\ell$ as small as possible in order to minimize the redundancy of the code. However, small values of $\ell$ can be problematic due to the following reasons: \begin{enumerate*}[label=(\roman*)] \item If $\ell$ is small, then the number of blocks given by $k/\ell$ is large. As a result, the number of cases to be checked by the decoder is large, and therefore this leads to an increase in the decoding complexity. \item Our theoretical analysis in Section~\ref{sec:8} shows that $\ell=\Omega(\log k)$ is required in order to guarantee an asymptotically vanishing probability of decoding failure (Remark~\ref{lw}). Hence, small values of $\ell$ do not guarantee a high probability of successful decoding. Furthermore, the theoretical analysis also shows that larger values of $\ell$ lead to a better upper bound on the probability of failure. \end{enumerate*}

Due to the reasons mentioned above, we specify $\ell$ based on~\eqref{eqell}. For $w=\Omega(\log k)$, we adopt the choice of $\ell=w$ since it gives a good trade-off between the code properties and simplifies the theoretical analysis. As for $w=o(\log k)$, we set $\ell=\log k$ since the choice of $\ell=w$ in this regime does not guarantee an asymptotically vanishing probability of decoding failure.

\begin{remark}
Since the redundancy is linear in $\ell$, then the condition of $\ell=\Omega(\log k)$ implies that the number of redundant bits should be at least logarithmic in order to correct the deletions successfully with high probability. This is in fact intuitive, because even for the case of one deletion, one would at least need to communicate the position of the deletion within the string, which would require a logarithmic number of redundant bits.
\end{remark}

\subsubsection{The minimum required number of MDS parities $c$} The decoding scheme for localized deletions requires $c\geq 3$ MDS parity symbols (2 for decoding the hypothetical block erasures and at least 1 for checking). However, the theoretical results in Theorems~\ref{thm:1} and~\ref{thm:2} suggest that $c\geq 4$ parity symbols are needed in order to guarantee an asymptotically vanishing probability of decoding failure. This discrepancy is due to the looseness of the theoretical upper bound on the probability of decoding failure for $c=3$. We discuss this in detail in Section~\ref{sec:6}, where we analyze the performance of GC codes for $c=3$ using numerical simulations.

\subsubsection{Comparison to the case of unrestricted deletions} The encoding and decoding schemes for localized deletions result in the code properties shown in Theorems~\ref{thm:1} and~\ref{thm:2}. Compared to the case of unrestricted deletions (Theorem~\ref{thm:5}), we observe the following: \begin{enumerate*}[label=(\roman*)] \item The construction presented for localized deletions enables correcting a number of deletions that {\em grows} with the message length $k$, as opposed to a {\em constant} number of unrestricted deletions. \item In the case of localized deletions, the parity bits can be recovered with zero-error without the use of a repetition code, which enhances the redundancy. \item The decoding complexity is quadratic for the case of localized deletions, as opposed to polynomial of degree~$\delta$ for unrestricted deletions.  \end{enumerate*} 

\section{Simulation Results}
\label{sec:6}
We tested the performance of GC codes for correcting deletions that are localized within a single window of size $w=\log k$ bits.  We performed numerical simulations for messages of length $k=128,256,512,1024,2048,$ and $4096$ bits. We also compared the resulting empirical probability of decoding failure to the upper bound in Theorem~\ref{thm:2}. 

\subsection{Results for $c=3$ MDS parity symbols}
Recall that the GC encoder (Fig.~\ref{fig:1}) adds $c>2$ MDS parity symbols resulting in $(c+1)\log k+1$ redundant bits when $\ell=w=\log k$. Here, we show the simulation results for the case of $c=3$ (Fig.~\ref{fig:sim1}), which corresponds to the minimum redundancy of GC codes.
\begin{figure}[h!]
\centering
\setlength\extrarowheight{1.2pt}
 \begin{tabular}{|c|c|c|c|c|}
\hline
\multirow{1}{*}{Config.} & \multicolumn{4}{c|}{$c=3$} \\ \cline{2-5} \hline
\multirow{2}{*}{$k$} & \multirow{2}{*}{$R$} & \multicolumn{3}{c|}{$Pr(F)$} \\   \cline{3-5} 
 & & $\delta=0.5w$ & $\delta=0.75w$ & $\delta=w$ \\ \hline
$128$  & $0.82$  & $9.06e^{-3}$ & $3.19e^{-2}$ & $4.19e^{-2}$\\ \hline
$256$ & $0.89$ & $5.06e^{-3}$ & $2.43e^{-2}$ & $4.11e^{-2}$\\ \hline
$512$  & $0.93$ & $3.81e^{-3}$ & $2.36e^{-2}$ & $3.96e^{-2}$\\ \hline
$1024$  & $0.96$ & $2.35e^{-3}$ & $2.34e^{-2}$ & $3.75e^{-2}$\\ \hline
$2048$  & $0.98$ & $2.09e^{-3}$ & $2.28e^{-2}$ & $3.59e^{-2}$\\ \hline
$4096$  & $0.99$ & $9.7e^{-4}$ & $1.31e^{-2}$ & $3.36e^{-2}$\\ \hline
\end{tabular}
\caption{\small The figures show the code rate $R=k/n$ and the empirical probability of decoding failure $Pr(F)$  of GC codes for different message lengths $k$ and number of deletions $\delta$. The $\delta$ deletions are localized within a window of size $w=\log k$ bits. The values of $R$ are rounded to the nearest $10^{-2}$. The results of $Pr(F)$ are averaged over $10^5$ runs of simulations. In each run, a message $\mathbf{u}$ chosen uniformly at random is encoded into the codeword $\mathbf{x}$. The positions of the window and the deletions in $\mathbf{x}$ are correspondingly chosen uniformly at random. The resulting string is then decoded.}
\label{fig:sim1}
\vspace{-0.5cm}
\end{figure}
\subsection{Results for $c=4$  and $c=5$ MDS parity symbols}
The table in Fig.~\ref{fig:sim2} shows the simulation results for $c=4$ with the same experimental setup described in Fig.~\ref{fig:sim1}. We also simulated the case of $c=5$ for the same values of $k$ and we were not able to detect any failure within $10^5$ runs of simulations.

\begin{figure}[h!]
\centering
\setlength\extrarowheight{1.2pt}
 \begin{tabular}{|C{0.8cm}|C{0.5cm}|c|c|c|c|}
\hline
\multirow{1}{*}{Config.} & \multicolumn{5}{c|}{$c=4$} \\ \cline{2-6} \hline
\multirow{2}{*}{$k$} & \multirow{2}{*}{$R$} & \multicolumn{4}{c|}{$Pr(F)$} \\   \cline{3-6} 
 & & $\delta=0.5w$ & $\delta=0.75w$ & $\delta=w$ & Bound\\ \hline
$128$  & $0.78$  & $7.0e^{-5}$ & $2.3e^{-4}$ & $2.7e^{-4}$ & $1.4e^{-1}$\\ \hline
$256$ & $0.86$ & $3.0e^{-5}$ & $6.0e^{-5}$ & $1.3e^{-4}$ &  $1.3e^{-1}$\\ \hline
$512$  & $0.92$ & $1.0e^{-5}$ & $3.0e^{-5}$ & $8.0e^{-5}$ & $1.1e^{-1}$\\ \hline
$1024$  & $0.95$ & $1.0e^{-5}$ & $3.0e^{-5}$ & $5.0e^{-5}$ & $1.0e^{-1}$\\ \hline
$2048$  & $0.97$ & $0$ & $0$ & $3.0e^{-5}$ & $9.1e^{-2}$ \\ \hline
$4096$  & $0.99$ & $0$ & $0$ & $0$ & $8.3e^{-2}$ \\ \hline
\end{tabular}
\caption{\small The values of the code rate $R$ and the empirical probability of failure $Pr(F)$ for different message lengths $k$ and number of localized deletions $\delta$, for the same experimental setup described in Fig.~\ref{fig:sim1}. The table also shows that GC codes perform better than what the theoretical upper bound in Theorem~\ref{thm:2} indicates.}
\label{fig:sim2}
\vspace{-0.5cm}
\end{figure}

\subsection{Comparison to the theoretical upper bound in Theorem~\ref{thm:2}}

\noindent To guarantee an asymptotically vanishing probability of decoding failure, the upper bound in Theorem~\ref{thm:2} requires that $c\geq 4$. In fact, the bound shows that the probability of decoding failure decreases logarithmically in $k$ for $c=4$, whereas for $c>4$, the probability decreases polynomially. Therefore, we make a distinction between the following three regimes. \begin{enumerate*}[label=(\roman*)] \item \mbox{$c=3:$}~Here, the theoretical upper bound is trivial, whereas in the simulation results in Fig.~\ref{fig:sim1} we observe that the probability of decoding failure is at most of the order of~$10^{-2}$. \item \mbox{$c=4$ (Fig.~\ref{fig:sim2}):}~The upper bound ranges from the order of $10^{-1}$ for $k=128$ to the order of $10^{-2}$ for $k=4096$, whereas the probability of decoding failure recorded in the simulations is at most of the order of $10^{-4}$ for $k=128$. Moreover, for $k=4096$, no failures were detected within $10^5$ runs of simulations. \item \mbox{$c> 4:$}~For $k=4096$, the upper bound is of the order of $10^{-5}$ for $c=5$, and of the order of $10^{-9}$ for $c=6$. In the simulations, no decoding failure was detected within $10^5$ runs for $c=5$. \end{enumerate*} In general, the simulations show that GC codes perform better than what the upper bound indicates. The looseness of the bound is due to the fact that the effect of Criterion 2 (Definition~\ref{def:1}) is neglected while deriving the bound (Section~\ref{sec:prf}). Furthermore, notice that for a fixed $k$, the empirical $Pr(F)$ increases as the number of deletions within the window increases. This dependence on $\delta$ is not reflected in the theoretical bound, since this bound is derived for the worst case of $\delta=w$, as we discuss in Section~\ref{sec:prf}. 

\section{Correcting Deletions Localized in Multiple Windows}
\label{sec:7}
In this section, we discuss how we generalize the previous results to the case where the deletions are localized in \mbox{$z>1$} windows. In terms of the encoding scheme, the first three encoding blocks remain the same as the ones for shown in Fig.~\ref{fig:1}, where the chunking length $\ell$ is specified based on~\eqref{eqell}.  As for the last encoding block (Block~IV, Fig.~\ref{fig:1}), we use a $(zw+1)$ repetition code to encode the parity bits. For decoding, the decoder first recovers the parity bits which are protected by the repetition code and then applies the guess and check method. 

The starting locations of the $z$ windows are distributed among the $k/\ell$ systematic blocks of the codeword. Furthermore, there are up to $zw$ bit deletions that are distributed among these $z$ windows. Therefore, the total number of cases to be checked by the decoder is 
\begin{equation*}
t=\mathcal{O}\left( \binom{k/\ell}{z} \binom{zw+z-1}{z} \right).
\end{equation*}
Recall that $z>1$ is a constant and $w\leq \ell$. Therefore,
\begin{equation}
\label{casez}
t= \mathcal{O}\left(\frac{k^z}{\ell^z} \cdot (zw)^z \right)= \mathcal{O}(k^z).
\end{equation}
Note that the exact value of $t$ depends on the sizes of the deletion windows. Nevertheless, as shown above, the order of $t$ is polynomial in $k$. The decoder goes over all these cases and applies the guess \& check method explained previously. The resulting code properties are given in Theorems~\ref{thm:3} and~\ref{thm:4} for $w=o(\log k)$ and $w=\Omega(\log k)$, respectively.
 
\begin{theorem}[Multiple windows with $w=o(\log k)$]
\label{thm:3} 
Guess \& Check (GC) codes can correct  in polynomial time  deletions that are localized within a constant number of windows $z>1$, where the size of each window is at most $w=o(\log k)$ bits, and the number of deletions in each window is at most $\delta\leq w$. The code has the following properties:
\begin{enumerate}[leftmargin=*]
\item Redundancy: $n-k = c(zw+1)\log k$ bits,
\item Encoding complexity is $\mathcal{O}(k \log k)$, and  decoding complexity is $\mathcal{O}\left(k^{z+1}\log k\right)$,
\item Probability of decoding failure: $Pr(F) = \mathcal{O}\left(k^{-(c-4z)} \right)$,
\end{enumerate}
where $c$ is a constant. 
\end{theorem}

\begin{theorem}[Multiple windows with $w=\Omega(\log k)$]
\label{thm:4} 
Guess \& Check (GC) codes can correct in polynomial time deletions that are localized within a constant number of windows $z>1$, where the size of each window is at most $w=\Omega(\log k)$ bits, and the number of deletions in each window is at most $\delta\leq w$. The code has the following properties:
\begin{enumerate}[leftmargin=*]
\item Redundancy: $n-k = c(zw+1)w$ bits,
\item Encoding complexity is $\mathcal{O}(k w)$, and  decoding complexity is $\mathcal{O}\left(k^{z+1}w\right)$,
\item Probability of decoding failure: $$Pr(F)=\mathcal{O}\left(k^z 2^{-w(c-3z)} \right),$$
\end{enumerate}
where $c$ is a constant. 
\end{theorem}

\begin{corollary}
\label{cor2}
For both regimes $w=o(\log k)$ and $w=\Omega(\log k)$ described in Theorems~\ref{thm:3} and \ref{thm:4}, the probability of decoding failure of GC codes for $z>1$ windows of localized deletions vanishes asymptotically in $k$ for $c\geq 4z$.
\end{corollary}

\noindent The proofs of Theorems~\ref{thm:3} and~\ref{thm:4} are provided in the Appendix. The proof of Corollary~2 follows directly from replacing the value of $c$ is the expression of the probability of decoding failure in Theorems~\ref{thm:3} and~\ref{thm:4} and is therefore omitted.

\begin{remark}
The results in Theorems~\ref{thm:3} and \ref{thm:4} are for a fixed number of windows $z>1$, where each window has a size at most $w$ that can grow with the message length~$k$. The decoding complexity in this case is polynomial of degree~$z+1$. Furthermore, Corollary~2 shows that the probability of decoding failure vanishes asymptotically if $c\geq 4z$.
\end{remark}


\section{Proof of Theorem~\ref{thm:1} and Theorem~\ref{thm:2}}
\label{sec:8}
Recall that the chunking length $\ell$ (Fig.~\ref{fig:1}) is specified based on the size of the window $w$, as given in~\eqref{eqell}. In this section, we provide the proofs for the redundancy, complexity, and probability of decoding failure of GC codes in terms of the parameter~$\ell$. Substituting $\ell=\log k$ and $\ell=w$ gives the results in Theorems~\ref{thm:1} and~\ref{thm:2}, respectively. The proofs provided in this section follow similar techniques as the ones used to prove the main result in~\cite{GC} (Theorem~\ref{thm:5}). The analysis requires some modifications to deal with the case of localized deletions. For the sake of completeness, we go over all the steps in detail.
\subsection{Redundancy}
The redundancy follows from the construction for the case of deletions that are localized within a single window (Fig.~\ref{fig:1}). The number of redundant bits is $c\ell +w+1$ (parities + buffer).
\subsection{Complexity}
The encoding complexity is dominated by the complexity of computing the $c$ MDS parity symbols. Computing one parity symbol involves $k/\ell$ multiplications of symbols in $GF(2^{\ell})$, and hence its complexity is $\mathcal{O}((k/\ell) \log^2(2^{\ell}))=\mathcal{O}(k\ell)$. Since $c$ is a constant, the overall encoding complexity is $\mathcal{O}(k\ell)$. The dominant factor in the decoding complexity is the part where the decoder goes over all the possible cases and applies erasure decoding for each case. Hence, the order of decoding complexity is given by the total number of cases multiplied by the complexity of erasure decoding. Recall that in each case only two erasures within the systematic symbols are to be decoded. These two erasures can be decoded by:  \begin{enumerate*}[label=(\roman*)] \item Multiplying the unerased systematic symbols by the corresponding encoding vectors and subtracting the obtained results from the corresponding parity symbols. \item Inverting a $2\times 2$ matrix. \end{enumerate*} Therefore, the overall decoding complexity is 
\begin{equation*}
\mathcal{O}\left(\left(\frac{k}{\ell}-1\right)\cdot k\ell \right)=\mathcal{O}\left(k^2\right).
\end{equation*}

\subsection{Probability of Decoding Failure}
\label{sec:prf}
The probability of decoding failure is computed over all possible $k-$bit messages. Recall that the message $\mathbf{u}$ is uniform iid, i.e., the bits of $\mathbf{u}$ are iid Bernoulli$(1/2)$. The message $\mathbf{u}$ is encoded as shown in Fig.~\ref{fig:1}. Consider $\delta\leq w$ deletions that are localized within a window of size at most $w$ bits. Due to the chunking length specified in~\eqref{eqell}, the $\delta$ deletions can affect at most two adjacent blocks in the codeword. Therefore, the decoder goes through $k/\ell -1$ cases, where in each case it assumes that certain two adjacent blocks were affected by the deletions. Let $\bm{\mathcal{Y}_i}$ be the random variable representing the $q$-ary string decoded in case \mbox{$i$, $i=1,2,\ldots,k/\ell-1$}. Let $\mathbf{Y} \in GF(q)^{k/\ell}$ be a realization of the random variable $\bm{\mathcal{Y}_i}$. We denote by $\mathcal{P}_r \in GF(q), r=1,2,\ldots,c,$ the random variable representing the $r^{th}$ MDS parity symbol, and let $p_r$ be a realization of the random variable $\mathcal{P}_r$. Also, let $\mathbf{G_r} \in GF(q)^{k/\ell}$ be the MDS encoding vector responsible for generating $\mathcal{P}_r$. For \mbox{$r=1,\ldots,c,$} we define the following random sets
\begin{align*}
\mathrm{A_r} &\triangleq \{ \mathbf{Y} \in GF(q)^{k/\ell} |~\mathbf{G_r^TY}= \mathcal{P}_r\}, \\ 
\mathrm{A_1^c} &\triangleq \mathrm{A_1} \cap \mathrm{A_2} \cap \ldots \cap \mathrm{A_c}.
\end{align*} 
For any set of given parities $p_1,\ldots, p_c$, $\mathrm{A_r}$ and $\mathrm{A_1^c}$ are affine subspaces of dimensions $k/\ell -1$ and $k/\ell-c$, respectively. Since $q=2^{\ell}$, we have 
\begin{equation}
\label{e:A}
\left \lvert \mathrm{A_r} \right \rvert = 2^{k-\ell} \text{ and} ~ \left \lvert \mathrm{A_1^c} \right \rvert = 2^{k-c\ell}.  
\end{equation}
$\bm{\mathcal{Y}_i}$ is obtained by decoding two erasures based on the first two parities, therefore, \mbox{$\bm{\mathcal{Y}_i}\in \mathrm{A_1}\cap \mathrm{A_2} $.} Note that $\bm{\mathcal{Y}_i}$ is not necessarily uniformly distributed over $\mathrm{A_1}\cap \mathrm{A_2}$. The next claim gives an upper bound on the probability mass function of  $\bm{\mathcal{Y}_i}$ for given arbitrary parities. 
\begin{claim} 
\label{claim:1}
For any case $i$, $i=1,2,\ldots,k/\ell -1$,
\begin{equation*}
Pr\left(\bm{\mathcal{Y}_i}=\mathbf{Y} | \mathcal{P}_1=p_1, \mathcal{P}_{2}=p_{2} \right) \leq \frac{1}{2^{k-3\ell}}.
\end{equation*}
\end{claim}
\noindent  We assume Claim~\ref{claim:1} is true for now and prove it shortly. Next, we use this claim to prove the upper bound on the probability of decoding failure. \\
\noindent In the general decoding scheme, there are two criteria which determine whether a case is {\em possible} or not (Definition~\ref{def:1}). Here, we upper bound $Pr(F)$ by taking into account Criterion~$1$ only. Based on Criterion~$1$, if a case $i$ is possible, then $\bm{\mathcal{Y}_i}$ satisfies all the $c$ MDS parities simultaneously, i.e., $\bm{\mathcal{Y}_i} \in \mathrm{A_1^c}$. Without loss of generality, we assume case~$1$ is the correct case. A decoding failure is declared if there exists a {\em possible} case~\mbox{$j$, $j=2,\ldots,k/\ell -1$}, that leads to a decoded string different than that of case~$1$. Namely, $\bm{\mathcal{Y}_j} \in \mathrm{A_1^c}$ and $\bm{\mathcal{Y}_j} \neq \bm{\mathcal{Y}_1}$. We first compute the conditional probability of failure for given parities $p_1,\ldots, p_c$. Note that in this case the set $\mathrm{A_1^c}$ is fixed. Let $t\triangleq k/\ell-1$, we have, 
\begin{align}
Pr\left(F | p_1\text{,...},p_c \right) &\leq Pr\left(\bigcup_{j=2}^{t}\{\bm{\mathcal{Y}_j}\in \mathrm{A_1^c},\bm{\mathcal{Y}_j} \neq \bm{\mathcal{Y}_1}\} \biggr\rvert p_1\text{,...},p_c \right) \\
&\leq \sum_{j=2}^{t} Pr\left(\bm{\mathcal{Y}_j}\in \mathrm{A_1^c},\bm{\mathcal{Y}_j} \neq \bm{\mathcal{Y}_1} |p_1,\ldots,p_c \right) \label{e:2} \\
&\leq \sum_{j=2}^{t} Pr\left(\bm{\mathcal{Y}_j}\in \mathrm{A_1^c} | p_1,p_2,\ldots,p_c \right) \label{e:4}\\
&= \sum_{j=2}^{t} \sum_{\mathbf{Y}\in \mathrm{A_1^c}} Pr\left(\bm{\mathcal{Y}_j}=\mathbf{Y} | p_1,p_2 \right) \label{e:new}\\
&\leq \sum_{j=2}^{t}  \sum_{\mathbf{Y}\in \mathrm{A_1^c}} \frac{1}{2^{k-3\ell}} \label{e:6}\\
&= \sum_{j=2}^{t}  \left \lvert \mathrm{A_1^c} \right \rvert \frac{1}{2^{k-3\ell}}  \\
&= \sum_{j=2}^{t}  \frac{2^{k-c\ell}}{2^{k-3\ell}} \label{e:8} \\
&=  \left(t-1\right)2^{-(c-3)\ell}\\
&= \left(\frac{k}{\ell}-2\right)2^{-(c-3)\ell} \label{e:10} \\
&\leq \frac{k }{\ell}\times  2^{-(c-3)\ell} \label{e:11}.
\end{align} 
\noindent \eqref{e:2} follows from applying the union bound. \eqref{e:4} follows from the fact that the conditional probability $Pr\left(\bm{\mathcal{Y}_j} \neq \bm{\mathcal{Y}_1}|\bm{\mathcal{Y}_j}\in \mathrm{A_1^c}, p_1,\ldots,p_c \right)\leq 1$. \eqref{e:new} follows from the fact that \mbox{$\bm{\mathcal{Y}_i}\in \mathrm{A_1}\cap \mathrm{A_2} $} and $\mathrm{A_1^c} \subset \mathrm{A_1}\cap \mathrm{A_2}$. \eqref{e:6} follows from Claim~\ref{claim:1}. \eqref{e:8} follows from \eqref{e:A}. \eqref{e:10} follows from the fact that \mbox{$t=k/\ell -1$}. Since \eqref{e:11} does not depend on the values of the parities, then the same bound is also obtained for $Pr(F)$. 
\begin{remark}
\label{lw}
Notice from~\eqref{e:11} that $Pr(F)$ goes to zero asymptotically in~$k$ if $\ell=\Omega(\log k)$. Hence, for a sub-logarithmic window size, i.e., \mbox{$w=o(\log k)$}, setting $\ell=w$ does not guarantee an asymptotically vanishing probability of decoding failure. To this end, we use $\ell=\log k$ for encoding when $w=o(\log k)$.
\end{remark}
\subsection{Proof of Claim~\ref{claim:1}}
Recall that $\bm{\mathcal{Y}_i}\in  \mathrm{A_1}\cap \mathrm{A_2}$ is the random variable representing the output of the decoder in case $i$. Claim~\ref{claim:1} gives an upper bound on the probability mass function of $\bm{\mathcal{Y}_i}$ for any $i$ and for given arbitrary parities $(p_1,p_{2})$. To find the bound in Claim~\ref{claim:1}, we focus on an arbitrary case $i$ ($i$ fixed) that assumes that the deletions have affected blocks $i$ and $i+1$. We observe $\bm{\mathcal{Y}_i}$ for all possible input $k-$bit messages, for a fixed deletion window\footnote{The window of $w$ bits, in which the $\delta$ deletions are localized, is fixed.}, and given parities $(p_1,p_{2})$. Hence, the observed case, the deletion window, and parities are fixed, while the input message varies. In this setting, we determine the maximum number of different inputs that can generate the same output. We call this number $\gamma$. Once we obtain $\gamma$ we can write
\begin{equation}
Pr\left(\bm{\mathcal{Y}_i}= \mathbf{Y} | W,  p_1,p_{2} \right) \leq \frac{\gamma}{\left \lvert  \mathrm{A_1}\cap \mathrm{A_2} \right \rvert} = \frac{\gamma}{2^{k-2\ell}}, \label{gamma}
\end{equation}
where $W$ is an arbitrary window of size $w$ bits in which the $\delta$ deletions are localized. We will explain our approach for determining $\gamma$ by going through an example for $k=32$ that can be generalized for any $k$. We denote by $b_o\in GF(2)$, $o=1,2,\ldots,k$, the bit of the message $\mathbf{u}$ in position $o$.

\begin{example}
\label{ex:5}
Let $k=32$ and \mbox{$\delta=w=\ell=\log k=5$}. Consider the binary message $\mathbf{u}$ given by
\begin{equation*}
\mathbf{u}=b_1~b_2~\ldots~b_{32}.
\end{equation*}
Its corresponding $q$-ary message $\mathbf{U}$ consists of $7$ symbols (blocks) of length $\log k=5$ bits each\footnote{The last symbol is padded to a length of 5 bits by adding zeros.}. The message $\mathbf{u}$ is encoded into a codeword $\mathbf{x}$ as shown in Fig.~\ref{fig:1}. We assume that the first parity is the sum of the systematic symbols and the encoding vector for the second parity is \mbox{$(1,\alpha,\alpha^2,\alpha^3,\alpha^4,\alpha^5,\alpha^6)$}~\footnote{The extension field used is $GF(32)$ and has a primitive element $\alpha$, with $\alpha^5=\alpha^2+1$.}. Suppose that the $\delta=w=5$ deleted bits in $\mathbf{x}$ are $b_3$ up to $b_7$. Recall that since $\ell=w$, the deletions can affect at most two blocks in $\mathbf{x}$. Suppose that the case considered by the decoder is the one that assumes that the deletions affected the $3^{rd}$ and $4^{th}$ block (wrong case). The decoder chunks the codeword accordingly, and symbols $3$ and $4$ are considered to be erased. The rest of the $q$-ary symbols are given by
\begin{align*}
S_1 &= \alpha^4 b_1+\alpha^3 b_2 +\alpha^2 b_8 + \alpha b_{9} + b_{10}, \\
S_2 &= \alpha^4 b_{11}+\alpha^3 b_{12} +\alpha^2 b_{13} + \alpha b_{14} + b_{15},\\
S_5 &= \alpha^4 b_{21}+\alpha^3 b_{22} +\alpha^2 b_{23} + \alpha b_{24} + b_{25}, \\
S_6 &= \alpha^4 b_{26}+\alpha^3 b_{27} +\alpha^2 b_{28} + \alpha b_{29} + b_{30}, \\
S_7 &= \alpha^4 b_{31}+\alpha^3 b_{32}.
\end{align*}
Notice that $S_1,S_2,S_5,S_6$ and $S_7$ are directly determined by the bits of $\mathbf{u}$ which are chunked at their corresponding positions. Hence, in order to obtain the same output, the bits of the inputs corresponding to these symbols cannot be different. For instance, if two messages differ in the first bit, then they will differ in $S_1$ when they are decoded, i.e., these two messages cannot generate the same output. Therefore, we refer to the bits corresponding to these symbols by the term ``fixed bits". The ``free bits", i.e., the bits which can differ in the input, are the $2w-\delta=5$ bits corresponding to the erasure $b_{16},b_{17},b_{18},b_{19},b_{20}$, in addition to the $\delta=5$ deleted bits $b_{3},b_{4},b_{5},b_{6},b_{7}$. The total number of ``free bits" is $2w=10$, so an immediate upper bound on $\gamma$ is $\gamma \leq 2^{10}$. However, these ``free bits" are actually constrained by the linear equations which generate the first two parities. By analyzing these constraints, one can obtain a tighter bound on $\gamma$. 

The constraints on the ``free bits" are given by the following system of two linear equations in $GF(32)$,
\begin{equation}
\label{edd1}
\left\{
\begin{array}{l}
\alpha^2 b_{3} + \alpha b_{4} + b_{5}+\alpha^4 b_{6}+\alpha^3 b_{7}  \\
~~~~~~~~~~~~+\alpha^4 b_{16}+\alpha^3 b_{17} + \alpha^2 b_{18} + \alpha b_{19} +b_{20} = p'_1,  \\
~~\\
\alpha^2 b_{3} + \alpha b_{4} + b_{5}+\alpha(\alpha^4 b_{6}+\alpha^3 b_{7}) \\
~~~~~~~+\alpha^3(\alpha^4 b_{16}+\alpha^3 b_{17} + \alpha^2 b_{18} + \alpha b_{19} +b_{20}) = p'_2,
\end{array}
\right.
\end{equation}
where $p'_1,p'_2\in GF(32)$ are obtained by the difference between the first and the second parity (respectively) and the part corresponding to the ``free bits". To upper bound $\gamma$, we upper bound the number of solutions of the system given by \eqref{edd1}. Equation~\eqref{edd1} can be written as follows
\begin{equation}
\label{edd3}
  \left\{
  \begin{array}{rrr}
    B_1 + B_2 + B_3  = p'_1, \\
   B_1 + \alpha B_2 + \alpha^3 B_3 = p'_2, 
  \end{array}
  \right.
\end{equation}
where $B_1, B_2$ and $B_3$ are three symbols in $GF(32)$ given by
\begin{align}
 B_1 &= \alpha^2 b_{3} + \alpha b_{4} + b_{5}, \label{p1}\\
B_2 &= \alpha^4 b_{6}+\alpha^3 b_{7}, \label{pp2} \\
B_3 &= \alpha^4 b_{16}+\alpha^3 b_{17} + \alpha^2 b_{18} + \alpha b_{19} +b_{20}. \label{p3}
\end{align} 
Notice that the coefficients of $B_1, B_2$ and $B_3$ in~\eqref{edd3} originate from the MDS encoding vectors. Hence, if we assume that $B_2$ is given, then the MDS property implies that \eqref{edd3} has a unique solution for $B_1$ and $B_3$. Moreover, since $B_1$ and $B_3$ have unique polynomial representations in $GF(32)$ of degree at most 4, then for given values of $B_1$ and $B_3$, \eqref{p1} and \eqref{p3} have at most one solution for $b_{3}, b_{4}, b_{5}, b_{16}, b_{17}, b_{18}, b_{19}$~and~$b_{20}$. Therefore, an upper bound on $\gamma$ is given by the number of possible choices of $B_2$, i.e., $\gamma \leq 2^2=4$.
\end{example}

The analysis in Example~\ref{ex:5} can be generalized for messages of any length $k$. Assume without loss of generality that $\delta=w$ deletions occur. Then, in general, the analysis yields $2w$~``free" bits and \mbox{$k-2w$}~``fixed" bits. Similar to~\eqref{edd3}, the \mbox{$2w$}~``free" bits are constrained by a system of $2$ linear equations in $GF(q)$. Note that this system of $2$ linear equations in $GF(q)$ does not necessarily have exactly $2$ variables. For instance, in Example~\ref{ex:5}, we had $3$ variables $(B_1,B_2,B_3)$ in the $2$ equations. This happens because of the shift caused by the deletions which could lead to the $2w$ ``free" spanning up to $3$ blocks, resulting in an additional symbol that is multiplied by a different MDS encoding coefficient. Therefore, since the difference between the number of symbols and number of equations is at most one, then the MDS property implies that the number of solutions of the system of equations is at most~\mbox{$2^{w}$}, i.e., $\gamma \leq 2^{w}$. Since $w\leq \ell$ from~\eqref{eqell}, the upper bound in~\eqref{gamma} becomes
\begin{equation}
Pr\left(\bm{\mathcal{Y}_i}= \mathbf{Y} | W,  p_1,p_{2} \right) \leq \frac{2^{\ell}}{2^{k-2\ell}}=\frac{1}{2^{k-3\ell}}. \label{c:1}
\end{equation}
The bound in~\eqref{c:1} does not depend on the locations of the localized deletions and holds for an arbitrary window position $W$. Therefore, the upper bound on the probability of decoding failure in~\eqref{e:11} holds for any window location that is chosen independently of the codeword. Moreover, for any given distribution on the window location (like the uniform distribution for example), we can apply the law of total probability and use the result from~\eqref{c:1} to get
\begin{equation}
Pr\left(\bm{\mathcal{Y}_i}= \mathbf{Y} | p_1,p_{2} \right) \leq \frac{1}{2^{k-3\ell}}. \label{ccc}
\end{equation}

\section{Conclusion}
In this paper, we introduced new explicit codes that can correct deletions that are localized within single or multiple windows in the codeword. These codes have several desirable properties such as: low redundancy, asymptotically optimal rate, efficient encoding and decoding, and low probability of decoding failure. We demonstrated these properties through our theoretical analysis and validated them through numerical simulations. Deriving fundamental limits for the problem of correcting localized deletions is an open problem and is one of the main future directions to consider. Another interesting direction is to apply these codes to file synchronization and compare their performance to other baseline algorithms such as rsync. 

\appendix[Proof of Theorem~\ref{thm:3} and Theorem~\ref{thm:4}]
The proof of Theorems~\ref{thm:3} and~\ref{thm:4} is a direct generalization of the proof of Theorems~\ref{thm:1} and~\ref{thm:2}. Next, we give the proof in terms of the chunking length~$\ell$. Substituting $\ell=\log k$ and $\ell=w$ gives the results in Theorems~\ref{thm:3} and~\ref{thm:4}, respectively. The redundancy follows from the construction (Section~\ref{sec:5}) where the $c\ell$ parity bits are repeated $(zw+1)$ times. Therefore, the redundancy is $c(zw+1)\ell$. The encoding complexity is $\mathcal{O}(k \ell)$, same as the case of a single window. The decoding complexity is given by the the product of the number of cases (guesses) and the complexity of decoding a constant number of systematic erasures (at most $2z$ erasures). Hence, the decoding complexity is \mbox{$\mathcal{O}(k^z\cdot k\ell)=\mathcal{O}(k^{z+1}\ell)$}. As for the probability of decoding failure, the statement of Claim~\ref{claim:1} can be generalized to $z>1$ windows as follows
\begin{equation}
Pr\left(\bm{\mathcal{Y}_i}=\mathbf{Y} | p_1,p_2,\ldots,p_{2z} \right) \leq \frac{2^{z\ell}}{2^{k-2z\ell}}=\frac{1}{2^{k-3z\ell}} \label{q:0}.
\end{equation}
The previous statement follows from the same steps of the proof of Claim~\ref{claim:1}, when applied to $2z$ equations, each corresponding to one MDS parity. Same as the proof of Theorems~\ref{thm:1} and~\ref{thm:2}, the rest of the proof of  follows from applying the union bound over the number of cases
\begin{align}
Pr(F|p_1,\ldots,p_{c})&\leq (t-1)\left \lvert \mathrm{A_1^c} \right  \rvert Pr\left(\bm{\mathcal{Y}_i}=\mathbf{Y} | p_1,\ldots,p_{2z}\right) \label{q:1}\\
&\leq (t-1)\cdot 2^{k-c\ell}\cdot  \frac{1}{2^{k-3z\ell}}  \label{q:2}\\
&= (t-1) 2^{-\ell(c-3z)} \label{q:3}\\
&= \mathcal{O}\left( k^z 2^{-\ell(c-3z)} \right) \label{q:4}.
\end{align} 
\eqref{q:1} follows from applying the union bound over the total number of cases $t$, similar to the proof of Theorems~\ref{thm:1} and~\ref{thm:2}. \eqref{q:2} follows from \eqref{e:A} and \eqref{q:0}. \eqref{q:4} follows from~\eqref{casez}. Since the bound in~\eqref{q:4} does not depend on the values of the parities, we obtain $Pr(F)=\mathcal{O}\left(k^z 2^{-\ell(c-3z)} \right)$. 
\bibliographystyle{ieeetr}

\bibliography{Refs}

\vspace{-7cm}
\begin{IEEEbiographynophoto}
{Serge Kas Hanna} (Member, IEEE) received the Diploma degree in computer and communications engineering from the Lebanese University, Faculty of Engineering, Roumieh, Lebanon in 2015, the M.S. degree in technology of information and communication systems from the Lebanese University, Doctoral school, Tripoli, Lebanon, in 2015, and the Ph.D. degree in electrical and computer engineering from Rutgers University, NJ, USA, in 2020. He is currently a Postdoctoral Researcher with
the Technical University of Munich, Germany. His research interests are in the broad area of coding theory, communication, and machine learning. 
\end{IEEEbiographynophoto}

\vspace{-7cm}
\begin{IEEEbiographynophoto}
{Salim El Rouayheb} (Member, IEEE) received the Diploma degree in
electrical engineering from the Faculty of Engineering, Lebanese University,
Roumieh, Lebanon, in 2002, the M.S. degree from the American University of
Beirut, Lebanon, in 2004, and the Ph.D. degree in electrical engineering from
Texas A\&M University, College Station, in 2009. He was a Post-Doctoral
Research Fellow with the University of California at Berkeley from 2010 to
2011 and a Research Scholar with Princeton University from 2012 to 2013.
He is currently an Associate Professor with the ECE Department, Rutgers
University, NJ, USA. His research interests are in information theory and
coding theory with a focus on applications to data security and privacy. He
was a recipient of the NSF Career Award and the Google Faculty Research Award.
\end{IEEEbiographynophoto}

\end{document}